\documentclass[]{elsart}
\usepackage{epsfig}
\usepackage{amssymb}
\usepackage{amsmath}

\begin{document}
\begin{frontmatter}
\title{Neutrinos from extragalactic cosmic ray interactions in the far infrared background}
\author[aaa]{E. V. Bugaev}
\author[bbb]{A. Misaki}
\author[ccc]{K. Mitsui}

\address[aaa]{Institute for Nuclear Research< 60th October Anniversary Prospect 7a, 117312 Moscow, Russia}
\address[bbb]{Advanced Research, Institute for Science and Engineering, Waseda University, 169-0092, Tokyo, Japan }
\address[ccc]{Faculty of Management Information, Yamanashi Gakuin University, Kofu 400, Japan}

\begin{abstract}
Extragalactic background of high energy neutrinos arising from
the interactions of cosmic ray protons with far-infrared extragalactic background 
radiation is calculated. The main assumption is that the cosmic ray spectrum at
energies higher than $10^8 GeV$ has extragalactic origin and, therefore, is proton
dominated. All calculations are performed with taking into account the possible 
cosmological evolution of extragalactic sources of cosmic ray protons as well
as infrared-luminous galaxies.
\end{abstract}

\end{frontmatter}

\section{Introduction}
\label{sec:intro}
The extragalactic backround of high energy neutrinos studied in this paper  arises from collisions of high energy cosmic ray (CR)
protons emitted by local sources (e.g., by active galactic nuclei) in extragalactic
space, with extragalactic photons. We will not consider here the more exotic
sources of extragalactic neutrino bachground (ENB) such as annihilations
of topological defects (e.g., cosmic strings) or decays of hypothetical relic
supermassive particles. So, ENB studied here began to be produced in relatively late
epochs (from cosmological point of view) of expansion when the galaxies of different
types and, in particular, the sources of CR's already existed. The first
calculations of ENB were performed about twenty years ago 
\cite{1,2,3}. Later, the detailed calculations of differential
spectra of ENB were done in works \cite{4,5,6,7,7a}. All these 
authors took into account the interactions of CR protons with
relic photons ($T\cong 2.7K$) only. 

Evidently, other components of the extragalactic radiation background also must
be taken into account in calculation of the extragalactic background of high
energy neutrinos. The interval of photon wavelengths which is potentially most important
from this point of view is $(1-1000)\mu m$, i.e., the infrared region. At last
fifteen years infrared astronomy developed very intensively (see, e.g.,
reviews of Franceschini \cite{8} and Hauser and Dwek \cite{9} ), and now the rather well-grounded
calculation of ENB from interactions of cosmic rays with extragalactic infrared photons 
became possible \cite{10,11}. 

Two main recent achievements of observational astronomy studying sources and
fluxes of infrared photons in extragalactic space are sufficient for the present work: 
i) the extremely infrared-luminous
galaxies have been discovered in the distant Universe (i.e., at substantial redshifts,
$z\sim 1$) and their brightness distribution were measured and ii) the intense isotropic
diffuse background radiation in the far-infrared, of extragalactic origin, have been 
revealed, and the spectral intensity of this background appeared to be of the 
same order as that of the optical extragalactic background.  The conclusion
from these observational facts is that some galaxies in the past should have been
much more bright in the far-infrared than in the optical. It means that a large fraction
of the energy emitted by distant galaxies should have been reprocessed by dust
at long wavelengths, and that violent processes of star formation took place in
the past in such massive systems.  This evidence of the strong cosmological evolution 
of infrared extragalactic sources is very important for theoretical predictions
of ENB as we show in this work. 

Technically, the problem of calculating the ENB from infrared
photons is rather complicate because now the photon number density
in extragalactic space, as a function of cosmological time, 
is not given (in the previous case of relic photons
this number is definitely known for all moments of time) and, therefore, must be 
calculated separately.

For the calculation of ENB one needs the following inputs.

1. {\it Photon energy spectra of infrared-luminous galaxies
and the number density of these galaxies in space.} Extensive information concerning
these characteristics is available in literature \cite{8,9}. For the concrete calculation
of ENB we chose the model of Beichman and Helou \cite{12} who derived the synthetic
spectral energy distributions in the large diapason of values of the total luminosities,
and the simple double power-law form of the local luminosity function (i.e., the
present-day number density) of infrared galaxies suggested by Soifer {\it et al} \cite{13}
and based on the IRAS data \cite{14}.

2. {\it The parameters characterizing the density and luminosity evolution of the 
infrared sources with look-back time}. Although the fact of a strong time evolution
of extragalactic infrared local sources is well grounded, it is reasonable,
however, at the present stage, to keep these parameters free.

3. {\it Extragalactic CR spectrum and intensity and the corresponding evolution parameters for
the sources of CR's.} We do not know well enough the extragalactic cosmic ray spectrum and,
therefore,  we are forced to use the crucial hypothesis about extragalactic origin of 
high energy CR's. Everywhere in our calculations we  use the extragalactic model (the
crossover energy $\sim 10^{17} eV$) and normalize our theoretical CR spectrum on 
experimental CR data.

Neutrino flux from interactions of high energy cosmic rays with infrared background 
was recently calculated by Stanev \cite{10}. Our work differs from that of \cite{10}
in several respects. The main difference is that in \cite{10} there is no non-trivial
cosmological evolution of infrared background: this evolution is assumed to be 
the same as of microwave background radiation (with $z_{max}=8$). Such an simplified
approach requires minimum information from infrared astronomy: only the spectral
intensity of the radiation at our epoch is needed, which can be, in principle,
taken directly from experimental data. Besides, in \cite{10} there is no separating
of contributions to ENB from infrared and optical diapasons of the background
while we calculate ENB from far-infrared part of the radiation  background
only (incidentally, it is well known that the time evolution of optical sources
is much weaker in comparison with infrared ones). 

The paper is organized as follows. In Sec.\ref{sec:2} we derive approximate
expression for the extragalactic CR spectrum for
nonzero redshifts using the continuous energy loss approximation. Sec.\ref{sec:3}
is dedicated to calculation of far infrared photon background and to obtaining the constraints
on parameters of cosmological evolution of this background.
In Sec.\ref{sec:4} we calculate the extragalactic  neutrino spectrum 
produced by interactions of cosmic ray protons with far infrared background,
using the inputs from Sec.\ref{sec:2} and Sec.\ref{sec:3}. 
The last section contains the result of the calculations.

\section{Extragalactic CR proton spectrum at different epochs}
\label{sec:2}

To obtain approximate expressions for the spectra and intensities 
of CR protons in extragalactic space we
use the cosmological transport (kinetic) equation without 
integral term, i.e., we work in the
continuous energy loss approximation introduced, for these problems,
by Berezinsky and Grigor'eva \cite{15}.
The validity of this approximation was studied in subsequent 
works, e.g., in the work of Yoshida and Teshima \cite{7}. It was shown
that this approximation works rather well and is not 
very good only for calculations of "local" cosmic ray spectra, i.e.,
cosmic ray spectra near the Earth from separate extragalactic sources
(in particular, the spikes near the GZK cut off predicted in calculations
using the continuous energy loss approximation are more 
sharply expressed than in exact Monte-Carlo calculations).
But in the case of diffuse cosmic ray spectra (i.e., the 
spectra integrated over redshift), the approximation of continuous
losses is satisfactory: in particular, it predicts correctly the characteristic 
dip-bump structure of the cosmic ray spectrum between $10^{19}$ 
and $10^{20}$ eV, at least in models with not very strong cosmological 
evolution of sources.

The cosmological transport equation for extragalactic CR protons is in our approximation
(see, e.g., the work of Blumental \cite{16})
\begin{equation}
\label{2.1}
\frac{\partial n (E,z)}{\partial z}+\frac{\partial}{\partial E}
\left[\beta(E,z)n(E,z)\right] - \frac{3n(E,z)}{1+z}+\frac{n(E,z)}{\lambda (E,z)}=g(E,z).
\end{equation}
Here, $n(E,z)$ is the differential number density of protons, having the 
dimension (unit volume$)^{-1}\cdot($unit of energy$)^{-1}$,i.e., $n(E,z)dE$
is the number of protons with a given redshift $z$ in the energy interval 
$dE$ near $E$. The function $\beta(E,z)$ is the change (loss) of proton
energy in unit interval of $z$, 
\begin{equation}
\label{2.2a}
\beta (E,z)=\frac{dE}{dz}=\frac{E}{1+z}-b(E,z)\frac{dt}{dz}.
\end{equation}
The first term in r.h.s. of Eq.(\ref{2.2a}) takes into account adiabatic
energy losses (those due to the cosmological expansion). The function
$b(E,z)$ describes the (non-adiabatic) proton energy losses per unit of time
spent in extragalactic space filled by photons of the diffuse radiation
background, and is defined by the expression
\begin{equation}
\label{2.2b}
t_p^{-1}(\gamma_p,z)=\frac{1}{z}b(E,z),
\end{equation}
where $t_p^{-1}(\gamma_p,z)$  is the cooling rate of protons with
gamma-factor $\gamma_p$ via $p\gamma\to \pi X$ and $p\gamma\to pe^{+}e^{-}$
reactions at the cosmological epoch with redshift $z$. Evidently, the cooling rate,
as a function of $z$, depends on the cosmological evolution of the radiation background.
Throughout this work we use the approximation that the relic photon gas (i.e., only
one component of the radiation background) is entirely responsible for the 
energy losses of cosmic ray protons in extragalactic space. In this 
approximation one has 
\begin{equation}
\label{2.2c}
t_p^{-1}(\gamma_p,z)= (1+z)^3t_p^{-1}(\gamma_p(1+z), 0).
\end{equation}
Using the text-book expression defining the energy loss coefficient $b(E)$,
\begin{equation}
\label{2.2d}
t_p^{-1}(\gamma_p,z)=-\frac{1}{E}\frac{dE}{dt}=-\frac{1}{E}b(E),
\end{equation}
one obtains the connection of the function $b(E,z)$, that enters Eq.(\ref{2.2a}), 
and $b(E)$. 
\begin{equation}
\label{2.2e}
b(E,z)=(1+z)^2 b(E(1+z)).
\end{equation}

The function $\lambda(E,z)$ in the kinetic equation is connected with 
the mean free path $\lambda(E,z)$ of a cosmic ray particle for an 
interaction leading to its absorbtion.
For simplicity, we neglect the corresponding term in the kinetic equation 
considering all proton interactions as non-catastrophic ones (in accordance
with our approximation of continuous energy losses).  Of course, some
absorbtion of cosmic ray protons in extragalactic space takes place, 
due to interactions with matter along the proton's path, and for taking
it into account the absorbtion term must be kept. 

The function $g(E,z)$ in r.h.s. of the kinetic equation describes the 
combined source of extragalactic cosmic rays. This source function 
can be written in the form 
\begin{equation}
\label{2.3}
g(E,z)=\rho(z)\eta(z)f(E)\frac{dt}{dz}.
\end{equation}

Here, $\rho(z)$ is the number density of local CR sources (e.g., AGNs) in the 
proper (physical) volume,
\begin{equation}
\label{2.4}
\rho(z)=\rho_0 (1+z)^3,
\end{equation}
$\eta(z)$ is the activity of each local source (the integrated number of produced
particles per second),
\begin{equation}
\label{2.5}
\eta(z)=(1+z)^{m}\eta_0\theta(z_{max}-z).
\end{equation}
Writing Eq.(\ref{2.5}) we assume that the cosmological evolution of cosmic 
ray sources can be parametrized by power law with the sharp 
cut-off at some epoch with redshift $z_{max}$
($m$ and $z_{max}$ are considered as parameters of a model of the combined source).
At last, the function $f(E)$ in Eq.(\ref{2.3}) describes a form of the differential 
energy spectrum of the local source
and has the simple normalization:
\begin{equation}
\label{2.6}
\int\limits_{E_0}^{\infty} f(E)dE=1.
\end{equation}
For the connection of differentials of cosmological time 
and redshift we use the simple formula
\begin{equation}
\label{2.7}
\frac{dt}{dz}=-\frac{1}{H_0(1+z)}\left[\Omega_m(1+z)^3+\Omega_{\Lambda}\right]^{-1/2}
\end{equation}
(assuming a flat Universe with matter and cosmological constant components). The 
matter density, in units of the critical density, 
$\Omega_m,$ is assumed to be equal $0.3$, and  $\Omega_{\Lambda}+\Omega_m=1$.

The solution of the kinetic equation is given by the expression 
\begin{equation}
\label{2.7a}
n(E,z)=\int\limits_{z_{max}}^{z}dx\exp\left\{\int\limits_x^z\left(\frac{3}{1+y}-
\frac{\partial\beta(z(y),y)}{\partial\xi(y)}\right)dy\right\}\cdot g(\xi(x),x),
\end{equation}
and the function $\xi(y)$ is a solution of the equation
\begin{equation}
\label{2.7b}
\frac{d\xi(y)}{dy}=\frac{\xi(y)}{1+y}+\frac{(1+y)b(\xi(y)\cdot(1+y))}{H_0\left[
\Omega_m(1+y)^3+\Omega_{\Lambda}\right]^{1/2}}
\end{equation}
with the initial condition
\begin{equation}
\label{2.7c}
\xi(z)=E.
\end{equation}
By differentiating the $\beta$ function given by Eq.(\ref{2.2a}) one obtains
\begin{equation}
\label{2.7d}
\frac{\partial\beta(\xi(y)\cdot y)}{\partial\xi(y)}=\frac{1}{1+y}+
\left.\frac{(1+y)^2}{H_0\left[\Omega_m(1+y)^3+\Omega_{\Lambda}\right]^{1/2}}b'(E)
\right|_{\xi(y)\cdot(1+y)},
\end{equation}
so, the  final formula for the solution is 
\begin{eqnarray}
\label{2.7e}
n(E,z)=\int\limits_{z_{max}}^zdx\left(\frac{1+z}{1+x}\right)^2\times\;\;\;\;\;\;\;\;\;\;\;\;\;\;\;\;\;\;\;\;\;\;\;\;\;\;\;\;\;\;\;\;\;\;\;\;\;\;\;\;\;\;\;\;\;\;\;\;\;\;\;\;\;\;\;  \nonumber  \\
\\
\times\exp\left\{\int\limits_z^x
\left[\left.\frac{(1+y)^2}{H_0\left[\Omega_m(1+y)^3+\Omega_{\Lambda}\right]^{1/2}}b'(E)
\right|_{\xi(y)\cdot(1+y)}\right]\right\}
\cdot g(\xi(x),x).\nonumber
\end{eqnarray}

Now, for the sake of the illustration, we consider the approximation which allows one
to obtain the analytical solution of the kinetic equation expressed through the simple
quadratures, which is rather good for approximate calculations and estimates. 

Namely, we approximate the energy loss coefficient $b(E)$ by the following simple expressions:
\begin{eqnarray}
\label{2.9}
b(E)=\beta_0 E \;\;\;;\;\;\; 3\cdot 10^9 GeV< E<10^{11}GeV,\nonumber\\
\\
b(E)=\beta_0' E\;\;\;\;\;\;\;\;\;\;\;\;\;\;\;\;;\;\;\;\;\;\;\;\;\;\;\;\;E>10^{11}GeV\nonumber
\end{eqnarray}
and $b(E)=0$ for $E<3\cdot 10^9 GeV$. Constants $\beta_0$ and $\beta_0'$ are:
\begin{equation}
\beta_0'=2\cdot 10^{-8}year ^{-1}\;\;\;;\;\;\;\beta_0=2.5\cdot 10^{-10} year ^{-1}.\nonumber
\end{equation}
We assume, further, the power law for the injection 
spectrum from the extragalactic local source,
\begin{equation}
\label{2.10}
f(E)=J_{inj}(E)=\frac{\gamma -1}{E_0}\left(\frac{E}{E_0}\right)^{-\gamma} (GeV^{-1}).
\end{equation}
($E_0$ is the cut-off value for the extragalactic part of the spectrum of the local
source; presumably, the particles with lower energies are somehow confined
inside the source). 

At last, we put in Eq.(\ref{2.7}) $\Omega_m=1$, $\Omega_{\Lambda}=0$. This 
approximation is not bad near $z=0$; even at $z=3$ it underestimates 
$dt/dz$ on a factor which is smaller than $2$. 

The resulting formulae for the cosmic ray proton spectrum appear to be 
different in three energy regions (everywhere below proton energy is substituted
in GeV):
\begin{eqnarray}
1.\;\;\;\;\;\;\;\;\;\;\;\;\;\;\;\;\;\;\;\;\;\;\;\;\;\;\;\;\;\; E<3\cdot 10^9 /(1+z)\nonumber\\
2.\;\;\;\;\;\; 3\cdot 10^9/(1+z)< E< 10^{11}/(1+z)\nonumber\\
3.\;\;\;\;\;\;\;\;\;\;\;\;\;\;\;\;\;\;\;\;\;\;\;\;\;\;\;\;\;\;\;\; E> 10^{11}/(1+z).\nonumber
\end{eqnarray}

In all final expressions for the spectrum there is the following common factor:
\begin{equation}
\label{2.11}
\Omega (E,z,\gamma ,E_0)=\frac{c}{4\pi}\cdot\frac{\rho_0\eta_0}{H_0}\cdot\frac{\gamma -1}{E_0}(1+z)^{\gamma +2}\cdot \left(\frac{E}{E_0}\right)^{-\gamma}.
\end{equation}

In the region 3. one obtains $(J(E,z)=\frac{c}{4\pi}n(E,z))$
\begin{equation}
\label{2.12}
J(E,z)=\Omega\int\limits_z^{z_{max}} dx (1+x)^{m-\frac{3}{2}-\gamma}
\exp\left\{178(1-\gamma)\left[(1+x)^{3/2}-(1+z)^{3/2}\right]\right\}.
\end{equation}
For the region 2. the formula for the spectrum is slightly more complicate:
\begin{equation}
\label{2.13}
J(E,z)=\Omega\int\limits_z^{min(z_{max}, x_b)}dx (1+x)^{m-\frac{3}{2}
-\gamma}\exp\left\{2.2\left[(1+x)^{3/2}-(1+z)^{3/2}\right]\right\}
\end{equation}
and the value $x_b$ is determined from the equation
\begin{equation}
\label{2.14}
\ln\frac{10^{11}(1+z)}{E(1+x_b)^2}=2.2\left[(1+x_b)^{3/2}-(1+z)^{3/2}\right].
\end{equation}
In the region 1. there are two subregions:

$1a.\;\;\;\;\; x_{b_1}\equiv\sqrt{\frac{3\cdot 10^9}{E}(1+z)} -1 > z_{max} \;\;\;;\;\;\; E> E_0;$
\begin{equation}
\label{2.15}
J(E,z)=\Omega \frac{1}{m-\frac{1}{2}-\gamma}\left[(1+z_{max})^{m-\frac{1}{2}-\gamma}-(1+z)^{m-\frac{1}{2}-\gamma}\right];
\end{equation}

$1b. \;\;\;\;\; x_{b_1}< z_{max}\;\;\;;$
\begin{eqnarray}
\label{2.16}
~
J(E,z)=\Omega \left[ \frac{1}{m-\frac{1}{2}-\gamma}
 \left[ (1+x_{b_{1}})^{m-\frac{1}{2}-\gamma}
-(1+z)^{m-\frac{1}{2}-\gamma} \right] \right. \nonumber \\
~
\\
~
\left. +\int\limits_{x_{b_1}}^{min(z_{max},x_{b_2})}
dx(1+x)^{m-\frac{3}{2}-\gamma}\exp \left\{ 2.2(1-\gamma)
\left[ (1+x)^{3/2}-(1+x_{b_1})^{3/2} \right] \right\} \right] \nonumber
\end{eqnarray}
and the value $x_{b_2}$ is determined from the equation similar to Eq.(\ref{2.14}):
\begin{equation}
\label{2.17}
\ln \frac{10^{11}(1+z)}{E(1+x_{b_2})^2}=2.2\left[(1+x_{b_2})^{3/2}-(1+x_{b_1})^{3/2}\right].
\end{equation}

Using expressions derived in this section one can calculate extragalactic cosmic
ray proton spectra for different cosmological epochs. The parameters that are
necessary for such a calculation are: $m,\;\;\; z_{max}, \;\;\; \gamma$ and
the product $(\rho_0\eta_0)$. Calculating $J(E,0)=\frac{c}{4\pi}\cdot n(E,0)$ 
with fixed values of  $m,\;\;\;z_{max},\;\;\;\gamma$ and normalizing it 
on the measured cosmic 
ray spectrum at energies $10^{17}-10^{19}$ GeV one can determine this 
product characterizing the 
combined cosmic ray source. As a result of this procedure 
we obtain the extragalactic cosmic 
ray proton spectrum for different epochs which can be used for the ENB calculation.

\section{Far infrared extragalactic background}
\label{sec:3}
For a calculation of the extragalactic radiation background it is convenient
to use the cosmological transport equation which is analogous to that used in
the previous section. The function which must be found is the number density 
of infrared photons at different cosmological epochs, $N^{IR}(E_{\gamma},z)$.
The expression for the source function of the kinetic equation is 
\begin{equation}
\label{3.1}
g^{IR}(E_{\gamma},z)=\int\frac{dL}{L}\rho(z,L)S^{IR}(E_{\gamma},L)\frac{1}{E_{\gamma}}\cdot\frac{dt}{dz}.
\end{equation}
Here, the function $S^{IR}$ is the spectral luminosity 
("spectral energy distribution") of  
a luminous local source (which is in this case an infrared-luminous 
galaxy with a given total infrared luminosity $L$). The normalization of this 
function is 
\begin{equation}
\label{3.3}
\int\limits_{E_{\gamma^{min}}}^{E_{\gamma}^{max}} S^{IR} (E_{\gamma},L)dE_{\gamma}=L,
\end{equation}
where, restricting oneself by the far infrared region, one takes 
\begin{equation}
\label{3.3a}
E_{\gamma}^{min}=1.243\cdot 10^{-3} eV\;\;\; (\lambda=10^3\mu m)\;\;\;;\;\;\; E_{\gamma}^{max}=1,243\cdot 10^{-1} eV\;\;\; (\lambda=10\mu m).
\end{equation}

The function $\rho(z,L)$ in Eq.(\ref{3.1}) is the number density of infrared-luminous 
galaxies with a given luminosity $L$ in the proper (physical) volume, which is connected
with the local luminosity function $\rho(0,L)$ by the relation
\begin{equation}
\label{3.4}
\rho(z,L)=\rho(0,\frac{L}{(1+z)^{\gamma_l}})(1+z)^{3+\gamma_{d}}.
\end{equation}
where 
$\gamma_d$ and $\gamma_l$ are parameters determining the cosmological evolution
of the luminosity of the local source, and the comoving density of these sources,
respectively.

For numerical calculation of the spectral energy distributions we use the model
of  Beichman and Helou \cite{12}.
These authors determined (theoretically) the synthetic spectral energy distributions
$F(E_{\gamma},L)$ for infrared-bright galaxies with total  far infrared luminosities
in the interval $(10^8-10^{11})\cdot L_{\odot}$, taking into account a number of 
components, the main of which are the "cirrus" component (infrared emission from a
galactic diffuse dust heated by the interstellar radiation field) and a hot 
starburst component (originating from the violent star formation processes
occuring inside dusty molecular clouds). The distributions $F(E_{\gamma},L)$
were calculated by Beichman and Helou in units of a spectral flux of energy
(in $Jy$), for a reference distance $1$ Mpc from  the galaxy. The connection between
$F(E_{\gamma}, L)$ and $S^{IR}(E_{\gamma},L)$ is 
\begin{equation}
\label{3.2}
S^{IR}(E_{\gamma},L)\cong 2\cdot 10^{53} F(E_{\gamma},L),
\end{equation}
and $S^{IR}(E_{\gamma},L)$ has dimension s$^{-1}$ (the luminosity
per unit interval of energy). 

For the local luminosity function $\rho(0,L)$ we use the analytic fit
of Soifer {\it et al} \cite{13} obtained from studies 
of IRAS infrared galaxies in the wavelength interval $12-100\mu m$,
\begin{eqnarray}
\label{3.5}
\rho(0,L)=10^{-2.07}\cdot L^{-2.31}\;\;\;,\;\;\; L\ge 10^{0,4}\nonumber\\
\\
\rho(0,L)=10^{-2.73}\cdot L^{-0.65}\;\;\;,\;\;\; L< 10^{0.4}.\nonumber
\end{eqnarray}

In Eq.(\ref{3.5}) the total luminosity L is measured in units of $(10^{10}L_{\odot})$ and 
$\rho(0,L)$ is in $(Mpc)^{-3}(\ln L)^{-1}$. The luminosity function was modelled 
in the interval
$$
10^{-1.8}< L < 10^{3},
$$
and was set to zero outside this range.

For a derivation of the final expression for the number density 
of infrared photons in extragalactic space we use the formula (\ref{2.7e}) of sec.\ref{sec:2},
putting there $g=g^{IR}$ and $b=0$ (i.e., we neglect energy losses 
of the infrared photons during their travelling in space). Substituting 
Eq.(\ref{3.1}) into Eq.(\ref{2.7e}) one obtains 
\begin{eqnarray}
\label{3.6}
N^{IR}(E_{\gamma},z)=\int\limits^{z_{max}}_{z} dx 
\left(\frac{1+z}{1+x}\right)^2
\int\frac{dL}{L}\rho(x,L)S^{IR}(\xi_{\gamma}(x),L)\times\nonumber\\
\\
\times\frac{1}{\xi_{\gamma}(x)}\cdot
\left[H_0(1+x)\sqrt{\Omega_m(1+x)^3+\Omega_{\Lambda}}\right]^{-1}.\nonumber
\end{eqnarray}
The function $\xi_{\gamma}(x)$ is a solution of the equation (compare with 
Eq.(\ref{2.7b}))
\begin{equation}
\label{3.6a}
\frac{d\xi_{\gamma}(x)}{dx}=\frac{\xi_{\gamma}(x)}{1+x}
\end{equation}
with the initial condition
\begin{equation}
\label{3.6b}
\xi_{\gamma}(z)=E_{\gamma},
\end{equation}
and, therefore, 
\begin{equation}
\label{3.6c}
\xi_{\gamma}(x)=E_{\gamma}\frac{1+x}{1+z}.
\end{equation}
The function $N^{IR}(E_{\gamma},z)$ is  measured in cm$^{-3}$eV$^{-1}$ if 
$E_{\gamma}$ is substituted in (eV), $\rho$ in cm$^{-3}(\ln L)^{-1}$ and
$S^{IR}$ in s$^{-1}$.

The expression (\ref{3.6}) is the main analytic result of this section. It allows
to calculate the extragalactic radiation background for any cosmological epoch
with taking into account the time evolution of the luminosity as well as 
density of the local sources. In the particular case $z=0$ one has from
Eq.(\ref{3.6}) the number density of infrared photons for our epoch.
Multiplying it on $E_{\gamma}$ and on a factor $\frac{c}{4\pi}$
one obtains the well known formula \cite{17} for the spectral flux
of the extragalactic background radiation,
\begin{equation}
\label{3.6e}
I(E_{\gamma})=\frac{c}{4\pi}\int \mbox{\bf L}(E_{\gamma} 
(1+z),z)\left|\frac{dt}{dz}\right|dz.
\end{equation}
Here, {\bf L}($E_{\gamma},z)$ is the spectral luminosity density 
of all luminous sources in a comoving volume element at redshift $z$,
\begin{equation}
\label{3.6f}
\mbox{\bf L}(E_{\gamma}(1+z))=\int dz\int d\ln L \rho_c(z,L)S^{^{IR}}(E_{\gamma}(1+z)),
\end{equation}
$\rho_c$ is the comoving luminosity function
\begin{equation}
\label{3.6g}
\rho_c(z,L)=\frac{\rho (z,L)}{(1+z)^3}.
\end{equation}
The function $E_{\gamma}I(E_{\gamma})$ is the "spectral intensity" of  the radiation
background (having dimension (energy$\cdot$cm$^{-2}\cdot$sec$^{-1}\cdot$ster$^{-1}$)),
which is commonly used for a presentation of data of background measurements.

As is mentioned in the Introduction, a distinctive property of infrared-luminous galaxies
is their very high rates of cosmological evolution, i.e., evolution with redshift $z$.
These rates sizeably exceed those measured for galaxies at other wavelengths. In particular, 
the comoving luminosity density from luminous infrared galaxies was at $z\sim 1$
on a factor about $10$ (or even more) larger than in the local Universe. Correspondingly,
the theoretical results which can be obtained for the extragalactic infrared radiation 
background are very sensitive to chosen values of the evolution parameters $\gamma_l$, $\gamma_d$.
From data on number counts of infrared galaxies at 15 and 24 $\mu m$ one can conclude
that there is some evidence of a strong density evolution as well as strong luminosity evolution
($\gamma_l\sim\gamma_d\sim 3$), although one cannot still exclude the possibility of an 
almost pure density evolution ($\gamma_d\gg \gamma_l$)

The detailed analysis of constraints on the evolution parameters following from data is
given in the review of Franceschini \cite{8}. It is argued in \cite{8} that the assumption 
of the very high evolution rates for the population of infrared sources (up to $\gamma_d \sim 8$
or more) is not in contradiction with data.The strong density evolution seems to be
very natural, at least for the population of starburst galaxies (those dominating at the
high luminosity end of the luminosity function) because the starburst phenomenon is caused,
most likely, by interactions and merging of galaxies (and the probability of such interactions
scales as square of the number density of galaxies in the physical volume).  

In the concrete numerical calculation we used the following walues of the evolution parameters
\begin{equation}
\label{3.6h}
\gamma_d=4,6,8\;\;\;\;\;;\;\;\;\;\;\gamma_l=0,\;\;\;\;\;;\;\;\;\;\;z_{max}=1,
\end{equation}
i.e., we assume that there is no luminosity evolution and that there is the
sharp cut-off after $z_{max}$. On Fig.1 we show the results of our 
far infrared background calculation for our epoch, together with experimental data,
taken from \cite{8}. We remind the reader, that we restrict ourselves in this work
by considering only the far-infrared part of the background ($10-1000 \mu m$). It is
seen from the figure that the choice $\gamma_d=6$ looks as the most appropriate. 
The slight shift of the maximum of the curves aside from the data is entirely due
to the features of theoretical spectral energy distributions used in the
calculation.

\begin{figure}[t!]
\label{fig:fig1}
\epsfig{file=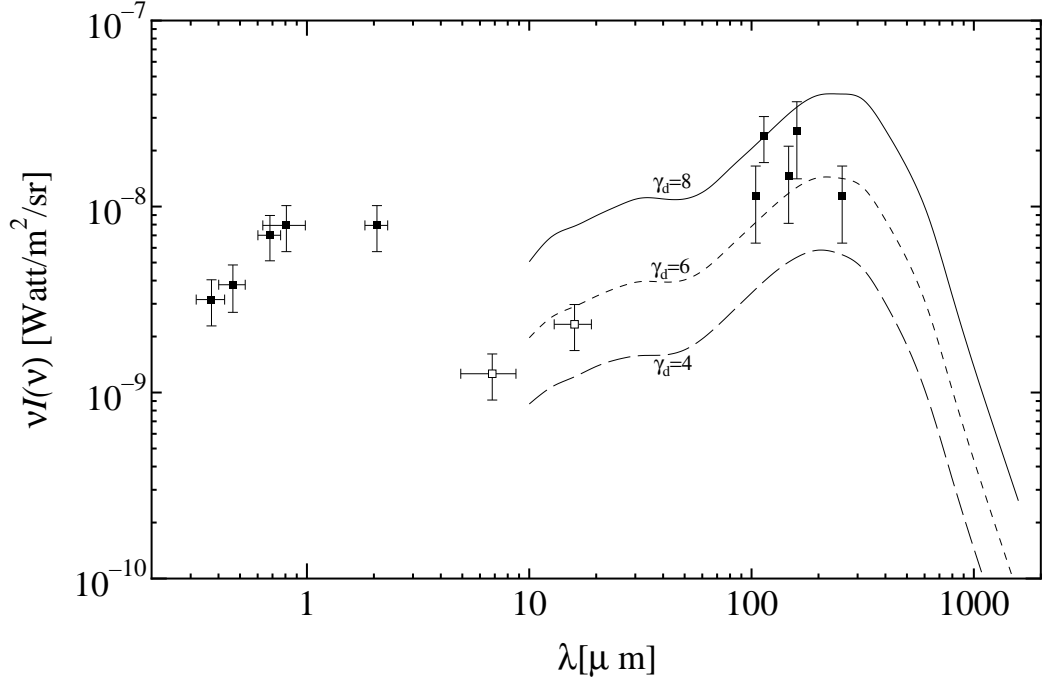,width=\columnwidth}
\caption{\footnotesize Extragalactic muon neutrino spectrum from interactions of cosmic rays 
with far infrared background radiation, for two values of density evolution
parameter $\gamma_d$.}
\end{figure}

One should note that our choice of the parameters $\gamma_d, z_{max}$ is the simplest one.
In the more realistic situation there is no sharp cut-off of the evolution:
it is most probable, from physical point of view, that the evolution rate changes
gradually.  Any possible variants can be considered modifying, trivially, Eq.(\ref{3.6}).

\section{Extragalactic spectra of high energy neutrinos}
\label{sec:4}
For simplicity we use in this work one-pion approximation, i.e., we assume that in 
$p\gamma$-reaction only two particles (neutron and pion) are produced. In this case,
as is well known (see, e.g., Hill and Schramm (1985)) the pions have approximately
isotropic distribution in the center of mass system and, as a consequence,
the step-like energy spectra in the observer system.

The main photoproduction reaction,
\begin{equation}
\label{4.1}
p_{CR}+\gamma_{infrared}\to \pi^{+}+n,
\end{equation}
and subsequent decays
\begin{equation}
\pi^{+}\to \mu^{+}+\nu_{\mu}, \;\;\;\;\;\;\;\mu^{+}\to e^{+}+\nu_e+\tilde \nu_{\mu}
\end{equation}
lead to production of $\nu_{\mu}$,$\tilde \nu_{\mu}$, $\nu_e$. In the present work we
will be interested only in ($\nu_{\mu}+\tilde\nu_{\mu}$)-flux. Therefore, we add together the
neutrino from pion and muon decays. Neutron produced in the photoproduction process will decay
giving proton, long before the next interaction with photons  of the background.

The number of proton-photon collisions per second and per units of photon energy and $\cos \theta_{\gamma}$ is equal to (for a given $z$)
\begin{equation}
\label{4.8}
\frac{dn_{coll}}{dt d E_{\gamma} d\cos\theta_{\gamma}}=\frac{c}{2}\sigma(s)
(1+\cos\theta_{\gamma})N^{IR}(E_{\gamma},z).
\end{equation}
Here, $\sigma(s)$ is the total cross section of the reaction (\ref{4.1}). Using the
connection between $E_{\gamma}$ and $E_{\gamma}^{lab}$, 
\begin{equation}
\label{4.6a}
E_{\gamma}^{lab}=E_{\gamma}(1+\cos\theta_{\gamma})\cdot \gamma_p,
\end{equation}
one obtains

\begin{equation}
\label{4.9}
\frac{dn_{coll}}{dt dE_{\gamma}^{lab}d\cos\theta_{\gamma}}=\frac{c}{2\gamma_p}
\sigma(E_{\gamma}^{lab})N^{IR}\left(\frac{E_{\gamma}^{lab}}{\gamma_p (1+\cos\theta_{\gamma})},z\right).
\end{equation}

Evidently, $n_{coll}$ is a function of $E_{p}$, $E_{\gamma}^{lab}$, $\cos \theta_{\gamma}$ and z.
Differential neutrino production spectrum from the pion and muon
decays is given by the following integral:
\begin{equation}
\label{4.10}
N_{\nu}^{prod}(E_{\nu},z)=\frac{4\pi}{c}\int dE_p\int dE_{\gamma}^{lab}\int d\cos\theta_{\gamma}
\frac{dn_{coll}}{dt dE_{\gamma}^{lab}d\cos\theta_{\gamma}}\cdot j_{\nu} (E_p,E_{\gamma}^{lab},E_{\nu})J(E_p,z),
\end{equation}
where $J(E_{p},z)$ is the extragalactic CR proton spectrum calculated in sec.\ref{sec:2}
and $j_{\nu}(E_p,E_{\gamma}^{lab},E_{\nu})$ is the neutrino spectrum per one $p\gamma$ collision.
The production spectrum is measured in ($cm^{-3}sec^{-1}GeV^{-1}$), and, as it
should be, it is the number of neutrinos produced in unit volume per second and per
unit interval of energy. So, this value has the same sense as the product $\rho (z)\eta (z)f (E)$ in 
the case of CR protons (see Eq.(\ref{2.3})).

The last step is the calculation of the neutrino extragalactic spectrum (or number density), 
i.e., the ENB,
by integration the $N_{\nu}^{prod}$ over all redshifts. The source function of the 
neutrino transport equation is 
\begin{equation}
\label{4.12}
g_{\nu}(E_{\nu},z)=N_{\nu}^{prod}(E_{\nu},z)\frac{dt}{dz}
\end{equation}
and the final result for the neutrino spectrum is
\begin{eqnarray}
\label{4.13}
J_{\nu}(E_{\nu})=\frac{c}{4\pi}\int\limits_{z_{max}}^0\frac{1}{(1+z)^2}
g_{\nu}(E_{\nu}(1+z),z)dz=\nonumber\\
\\
=\frac{c}{4\pi H_0}\int\limits_0^{z_{max}}
\frac{1}{(1+z)\left[\Omega_m(1+z)^3+ \Omega_{\Lambda}\right]^{1/2}} 
N_{\nu}^{prod}(E_{\nu}(1+z),z)dz.\nonumber
\end{eqnarray}

\section{Results and discussions}
\label{sec:5}
As is noted in the Introduction, we used in this work the basic assumption that 
ultra high energy cosmic rays have extragalactic origin. It means that at $E>E_{0}$,
where $E_0$  is some crossover energy, all cosmic rays detected near the Earth 
came from extragalactic space where they had interactions with extragalactic 
radiation background and, moreover, they are mostly protons. This hypothesis allows us to normalize our theoretical 
cosmic ray spectrum obtained in sec\ref{sec:2} on data of cosmic ray measurements 
at energies $> E_0$.

We did not attempt in this work to sew accurately the galactic and extragalactic spectra
near crossover energy, it is the separate and delicate problem. It is enough for us, 
at this stage, that the characteristic features of the experimental cosmic ray 
spectrum at $10^{18}-10^{19}$ eV, i.e., the dip-bump structure and the beginning of the
GZK cut-off, can be well described in our model, by the proper choice of the
parameters $\gamma, m, z_{max}$. In general, the possibility of a describing 
of the cosmic ray spectrum at $E\sim 10^{18}-10^{19}$ eV by models with
extragalactic origin of high energy cosmic rays was proved in a 
series of papers by Berezinsky with coauthors \cite{15,18}. 

We chose for the crossover energy the value $E_0=10^{17}$ eV. The final 
results do not depend very much on the concrete choice of $E_0$ if $E_0$
is much lower than the threshold of the photoproduction on infrared photons,
\begin{equation}
\label{5.0}
E_{thr}\sim \frac{m_pm_{\pi}}{\gamma_{far-infr}}\sim 10^{19} eV.
\end{equation}

The experimental form of the cosmic ray spectrum at $10^{18}-10^{19}$ eV
is well described with the following set of parameters: 
\begin{equation}
\label{5.0a}
\gamma=2.5\;\;\;\;,\;\;\; m=3.5\;\;\;\;\;\;,\;\;\; z_{max}=5.
\end{equation}
We see that the value of $z_{max}$ characterizing the evolution 
of cosmic ray sources is much higher than the corresponding value
chosen in sec.\ref{sec:3} for infrared sources. It is not the contradiction
because the physical nature of sources is clearly different in these 
two cases: the contributions of AGNs to the infrared background is rather small
(see, e.g., \cite{19}).

Normalizing the theoretical cosmic ray spectrum at $z=0$ with the chosen set 
of parameters on cosmic ray data one obtains 
\begin{equation}
\label{5.0b}
\rho_0\eta_0\cong 10^{-41} (cm^{-3}sec^{-1}).
\end{equation}
This value corresponds to the total luminosity required for the production
of cosmic rays with $E>E_0$
\begin{equation}
\label{5.0c}
\sim 4.5\cdot 10^{45} (erg Mpc^{-3} Yr^{-1}).
\end{equation}
If, e.g., there is one cosmic ray source per volume $10^6 (Mpc)^3$, the activity of 
each such source is about $10^{44}erg/s$ (for $E>10^{17}$ eV).

\begin{figure}[t!]
\label{fig:fig2}
\epsfig{file=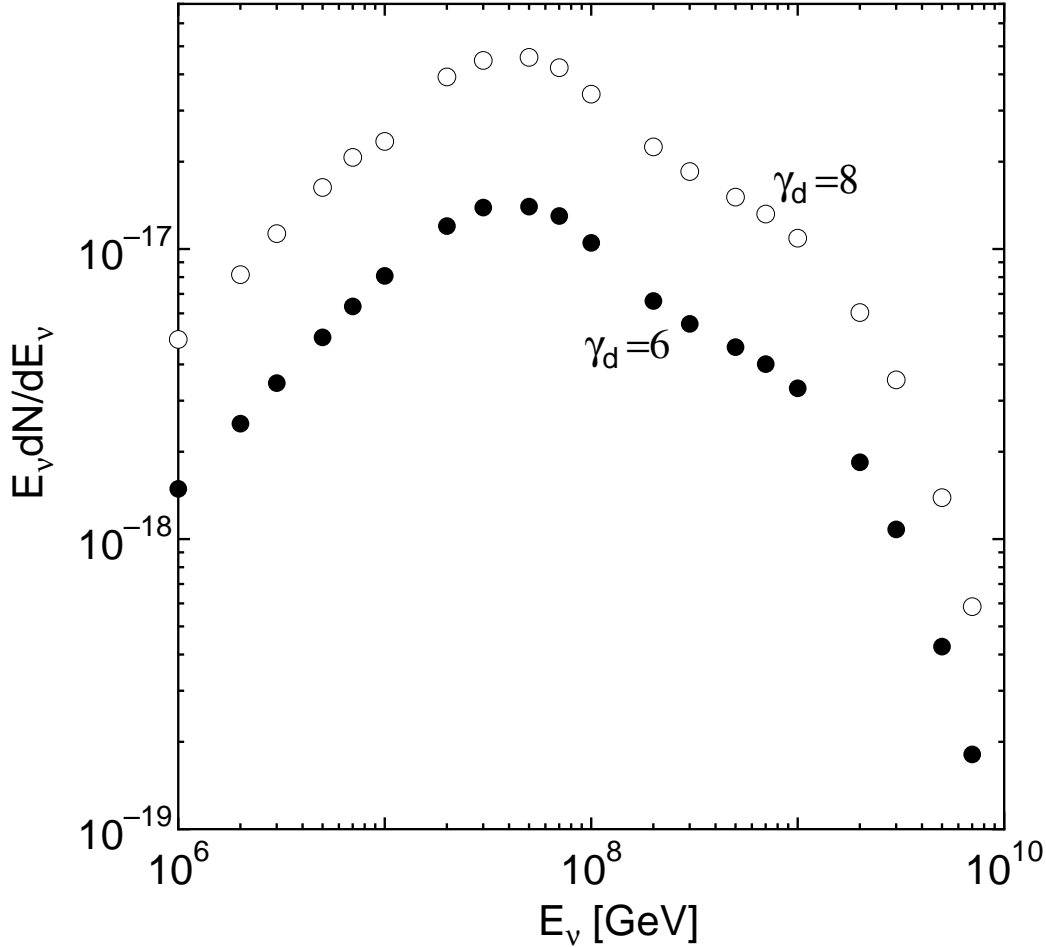,width=\columnwidth}
\caption{\footnotesize Far infrared radiation background calculated using the model 
of Beichman and Helou, for three values of the density evolution parameter
$\gamma_d$.}
\end{figure}

The resulting curves of the muon neutrino extragalactic background from 
interactions of cosmic rays with far-infrared photons of the radiation 
background are shown, for two chosen values of the parameter $\gamma_d$,
on Fig.2. One can see a strong dependence of the result on the 
assumed evolution law of the infrared background. The order of magnitude of ENB
is qualitatively the same as in the Stanev's work \cite{10}. In spite of the 
fact that the density of infrared photons in extragalactic space is much
smaller ($\sim 1.5$ photons$/$cm$^3$) than that of relic microwave photons,
the neutrino background appears to be not so small, due to the lower 
threshold for photoproduction and, last but not least, due to much
stronger time evolution of infrared background in comparison with 
that of relic photons. The predicted neutrino fluxes are comparable, 
more or less, with the neutrino fluxes from other extragalactic sources
at energy region near $10^{17}$ eV ($\gamma$-ray bursts, topological defects, etc)
and deserve further theoretical and, in future, experimental studies.

\end{document}